\documentclass{sf2a-conf2015}
\usepackage{graphicx}
\usepackage[colorlinks=true]{hyperref}
\usepackage[]{natbib}  
\usepackage{epstopdf}

\def\BibTeX{{\rm B\kern-.05em{\sc i\kern-.025em b}\kern-.08em
    T\kern-.1667em\lower.7ex\hbox{E}\kern-.125emX}}
\bibpunct{(}{)}{;}{a}{}{,}  %%%%%%%%%%%%%  A&A bibliography style
\usepackage{bm}
\usepackage{wasysym}
\newcommand*{\doi}[2]{\href{http://dx.doi.org/#1}{#2}}

\begin{document}

\TitreGlobal{SF2A 2015}
%%-----------------------------------------------------------------
%%      the top matter
%%

\title{Tests of gravitation with GAIA observations of Solar System Objects}

\runningtitle{Tests of GR with Gaia}

\author{A. Hees}\address{Department of Mathematics, Rhodes University, Grahamstown 6140, South Africa. a.hees@ru.ac.za}

\author{D. Hestroffer}\address{IMCCE, Observatoire de Paris - PSL Research University, UPMC Universit\'es P06, Universit\'e Lille 1, CNRS, France}

%% IF Author3 has the same affiliation than Author1:
\author{C. Le Poncin-Lafitte}\address{SYRTE, Observatoire de Paris, PSL Research University, CNRS, Sorbonne Universit\'es, UPMC Univ. Paris 06, LNE, 61 avenue de l'Observatoire, 75014 Paris, France}
\author{P. David$^2$}

%% IF Author3 has its own affiliation:
%\author{C.\,E. Author3}\address{Dept. of Chess, University of Games, 35101 Las Vegas, Monaco} 

%% IF Author3 has two affiliations, the one of Author1 and a second one:

%% Keep this line, even if the page will be settled afterwards.
\setcounter{page}{237}

%%-----------------------------------------------------------------

\maketitle

%%-----------------------------------------------------------------
%%        The abstract
%% 
%%  Warning!  within the abstract:
%%  - do not use macros. 
%%  - do not use commands like: \cite, \citet, \citep ... etc.

\begin{abstract}
In this communication, we show how asteroids observations from the Gaia mission can be used to perform local tests of General Relativity (GR). This ESA mission, launched in December 2013, will observe --in addition to the stars-- a large number of small Solar System Objects (SSOs) with unprecedented astrometric precision. Indeed, it is expected that about 360,000 asteroids will be observed with a nominal sub-mas precision.

Here, we show how these observations can be used to constrain some extensions to General Relativity. We present results of SSOs simulations that take into account the time sequences over 5 years and geometry of the observations that are particular to Gaia. We present a sensitivity study on various GR extensions and dynamical parameters including: the Sun quadrupolar moment $J_2$, the parametrized post-Newtonian parameter $\beta$, the Nordtvedt parameter $\eta$, the fifth force formalism, the Lense-Thirring effect, a temporal variation of the gravitational parameter $GM_{\astrosun}$ (a linear variation as well as a periodic variation), the Standard Model Extension formalism, \dots \, Some implications for planetary ephemerides analysis are also briefly discussed.
\end{abstract}

%% Insert the keywords (to appear in the ADS indexing)
%% Keywords must be separated by a comma

%%-----------------------------------------------------------------

\section{Introduction}
%%---------------------
This year marks the centenary of the publication of the classical theory of General Relativity (GR), the current paradigm to describe the gravitational interaction. Since its publication in 1915, GR has been confirmed by experimental observations. Although very successful so far, it is nowadays commonly admitted that GR is not the ultimate theory of gravitation. Attempts to develop a quantum theory of gravitation or to unify gravitation with the others fundamental interactions lead to deviation from GR. Moreover, observations requiring the introduction of Dark Matter and Dark Energy are sometimes interpreted as a hint that gravitation presents some deviations from GR at large scales.

GR is built upon two fundamental principles. The first principle is the Einstein Equivalence Principle (EEP) which gives to gravitation a geometric nature. More precisely, EEP implies that gravitation can be identified to space-time curvature which is mathematically described by a space-time metric $g_{\mu\nu}$. If the EEP postulates the existence of a metric, the second building of GR specifies the form of this metric. The metric tensor is determined by solving the Einstein field equations which can be derived from the Einstein-Hilbert action.

The EEP has been thoroughly tested~\citep[][and references therein]{will:2014la} by considering Universality of Free Fall experiments \citep{schlamminger:2008zr,adelberger:2009fk,williams:2009ys,muller:2012ys}, constancy of the constants of Nature \citep{uzan:2011vn,rosenband:2008fk,guena:2012ys,minazzoli:2014xz,hees:2014uq}, redshift experiments \citep{vessot:1980fk,delva:2015fk}, anisotropy in the velocity of light \citep{will:2014la}, \dots\, 

On the other hand, in the last 50 years, mainly two formalisms have been used to test the form of the metric: the parametrized post-Newtonian (PPN) formalism fully described in \citet{will:1993fk} and the fifth force formalism described in \citet{talmadge:1988uq,adelberger:2009fk}. The PPN formalism is making a nice interface between theoretical developments and observations by parametrizing deviations from GR at the level of the space-time metric by means of 10 dimensionless coefficients. The fifth force formalism considers a modification of the Newtonian potential of the form of a Yukawa potential. The parameters entering these formalisms have been severely constrained by diverse measurements \citep[for a review, see][]{will:2014la}: spacecraft tracking \citep{bertotti:2003uq,konopliv:2011dq}, Lunar Laser Ranging \citep{williams:2009ys,muller:2012ys}, planetary ephemerides \citep{fienga:2011qf,pitjeva:2013fk,verma:2014jk,fienga:2015rm}, Very Long Baseline Interferometry \citep{lambert:2009bh}, etc\dots \, Despite very stringent constraints, we have strong theoretical motivations to pursue these tests and to consider formalisms beyond the standard PPN and fifth force formalism \citep[for some motivations, see for example][]{hees:2012fk,hees:2013vn,hees:2014jk}.

 Launched in December 2013, the ESA Gaia mission is scanning regularly the whole celestial sphere once every 6 months providing high precision astrometric data for a huge number ($\approx$ 1 billion) of celestial bodies. In addition to stars, it is also observing SSOs, in particular asteroids. One can estimate that about 360 000 asteroids will be regularly observed. The high precision astrometry (at sub-mas level) will allow us to perform competitive tests of gravitation and to provide new constraints on alternative theories of gravitation. These constraints will be complementary to the ones existing currently since relying on different bodies, on different type of observations and therefore sensitive to other systematics.

In this communication, we report the first results of a sensitivity study of Gaia SSOs observations to several modifications of the gravitation theory. Correlations between the different dynamical parameters (in particular with the Sun quadrupolar moment) are also assessed.

\section{Simulations and parameters estimation}
We have simulated the trajectories of asteroids using the standard post-Newtonian equations of motion in a heliocentric frame. The mutual interactions between the asteroids are neglected, the Sun oblateness parameter $J_2$ is considered and the perturbations from the different planets and the Moon are modelled by using INPOP10e ephemerides~\citep{fienga:2013yg}. The initial conditions used for the simulations are provided by the ASTORB database\footnote{see \url{http://www.naic.edu/~nolan/astorb.html}} and a match between the asteroids trajectories with the Gaia scanning law is performed to find the observation times for each asteroid. Simultaneously with the equations of motion, we integrate the variational equations (for a detailed presentation of the method, see~\citet{hestroffer:2010qv,mouret:2011qf,mouret:2011uq}). The simulated asteroid trajectories are transformed into astrometric observables as well as their partial derivatives with respect to the parameters considered in the covariance analysis. The parameters considered here are twofolds: (i) local parameters specific to each asteroid (e.g. their 6 initial conditions) and (ii) global parameters that are common for all asteroids (the Sun $J_2$ and the parameters that characterize the gravitation theory). Our sensitivity study is performed by computing the covariance matrix to assess the correlations between the estimated parameters and the formal uncertainties reachable using Gaia observations. In this proceedings, the simulations performed include only 10,000 asteroids and the astrometric accuracy used is 0.2 mas. Full simulations of 360,000 asteroids with updated and better precision are ongoing and will be published in the future. 

\section{Sensitivity study to various global parameters}
In all the following, we consider modifications of the gravitation theory that do not produce any important effects on the light propagation. Therefore, we concentrate on the orbital dynamics only and neglect any modification in the propagation of the light. Simulations using the Time Transfer Formalism~\citep{teyssandier:2008nx,hees:2012fk,hees:2014fk,hees:2014nr} have been performed to ensure that the effects of the considered gravitation modifications on the light propagation can safely be neglected. 

In the following subsections, we report a sensitivity study performed by a global inversion that includes the 6 initial conditions for each of the 10,000 asteroids, the Sun $J_2$ and the parameters characterizing the gravitation theory.

\subsection{Sun quadrupolar moment, PPN parameter and Nordtvedt effect}
First of all, we consider the standard PPN parameter $\beta$ \citep{will:1993fk,will:2014la}. It is well known that this parameter is highly correlated with the Sun quadrupole moment $J_2$. Indeed, the secular advances of the perihelia produced by these two parameters are given by
\begin{equation}
	\left<\frac{d\omega}{dt}\right>=(2+2\gamma-\beta)n\frac{GM}{c^2a(1-e^2)}+\frac{3}{2}n\frac{J_2R_e^2}{a^2(1-e^2)^2}\, ,
\end{equation}
where $\omega$ is the argument of the perihelia, $GM$ is the Sun gravitational parameter, $c$ the speed of light, $\gamma$ and $\beta$ are the PPN parameters, $J_2$ the Sun quadrupole moment, $R_e$ the Sun equatorial radius, $a$ the semimajor axis of the orbit considered, $e$ its eccentricity and $n=(GM/a^3)^{1/2}$ the mean motion. In this analysis, we use $\gamma=1$ since this parameter is better determined by other types of observations like the Shapiro time delay~\citep{bertotti:2003uq} and by Gaia itself that will be able to constrain it at the level of $10^{-6}$ by observing light deflection~\citep{mignard:2010kx}.

The fact that a large number of asteroids are considered with various different orbital parameters helps to decorrelate the two parameters. In this simulation, we obtain a sensitivity around $\sigma_{J_2}\sim 10^{-7}$ and $\sigma_\beta\sim 7\times 10^{-4}$ with a correlation coefficient of 0.56 between the two parameters. These sensitivities are 1 order of magnitude lower than the ones obtained with planetary ephemerides~\citep{pitjeva:2014fj,verma:2014jk,fienga:2015rm}. The consideration of the complete set of asteroids (360,000 instead of 10,000) will improve slightly our current estimation. Planetary ephemerides will likely remain more powerful but Gaia will provide an independent and complementary estimation.

In addition to these two parameters, we also consider a violation of the Strong Equivalence Principle (SEP). Such a violation appears in all alternative theories of gravitation. One effect produced by a violation of the SEP is that the trajectories of self-gravitating bodies depend on their gravitational self energy $\Omega$ (violation of the universality of free fall). It is characterized by a difference between the gravitational and the inertial mass usually parametrized by the Nordtvedt parameter $\eta$
\begin{equation}
	m_g=m_i+\eta\frac{\Omega}{c^2} \, ,
\end{equation}
where $m_g$ is the gravitational mass and $m_i$ is the inertial mass. Using the same modelling as in \citep{mouret:2011uq}, the estimated uncertainty on $\eta$ using simulations of 10,000 asteroids is $9\times 10^{-4}$. This parameter does not change the estimations on the $\beta$ and $J_2$ parameters and is not correlated to them. The only constraint available currently on $\eta$ is provided by Lunar Laser Ranging measurements and is at the level of $4.5 \times 10^{-4}$~\citep{williams:2009ys}. 

Moreover, in the PPN framework, the $\eta$ parameter is unambiguously related to the PPN parameters~\citep{will:1993fk} by the relation
\begin{equation}
	\eta=4\beta-\gamma-3 \, .
\end{equation}
Instead of estimating 3 independent parameters ($J_2$, $\beta$ and $\eta$), one can introduce the last relation and estimate only two of them ($J_2$ and $\beta$). By doing so, the estimated uncertainty on $J_2$ and $\beta$  becomes $\sigma_{J_2}\sim 9\times 10^{-8}$ and $\sigma_{\beta}\sim 2\times 10^{-4}$. Considering a violation of the SEP predicted by the PPN framework leads to an improvement in the estimation of the $\beta$ PPN parameter. Moreover the correlation coefficient between $\beta$ and $J_2$ drops from 0.56 to 0.18, which can be a considerable improvement. Similar conclusion holds for planetary ephemerides analysis and this gives a strong motivation to consider violation of the SEP with planetary ephemerides gravitation tests. 

\subsection{Lense-Thirring effect}
The Lense-Thirring effect is a purely relativistic frame-dragging effect produced by the angular momentum of bodies. While the Lense-Thirring effect from the Earth has been detected with the LAGEOS satellite~\citep{ciufolini:2004uq} (see nevertheless the controversy raised in~\citet{iorio:2011ys}) and with the Gravity Probe B mission~\citep{everitt:2011fk}. The Sun Lense-Thirring has never been directly detected so far. Such a direct measurement would provide a new way to estimate the Sun angular momentum, an important quantity to assess interior models of the Sun~\citep{pijpers:2006kx}. Possibilities to measure the Sun Lense-Thirring effect has been mentioned in \citep{iorio:2011ys,iorio:2011zr,iorio:2012fr,iorio:2012uq}. Nevertheless, as mentioned in~\citep{folkner:2014uq}, planetary ephemerides analysis does not allow to disentangle the Sun Lense-Thirring from the Sun $J_2$ (even with the latest Messenger data). The two effects are completely correlated. This is partially due to the fact that all planets are orbiting in a very similar plane with a nearly circular orbit. Considering the asteroids can help since they provide a wide range of different orbital parameters especially including larger inclinations that may reduce this correlation. With the 10,000 asteroids considered in this study, the uncertainty on the Sun Lense-Thirring is a factor 5 larger than its actual value. This means that a direct detection of this effect with Gaia observations seems difficult. This conclusion should be reassessed by the consideration of the full set of asteroids. Moreover, the combination of the Gaia dataset with radar observations performed at UCLA~\citep{margot:2010fk} may also improve this conclusion. Finally, it would be interesting to assess the gain that an analysis combining planetary ephemerides with Gaia observations would bring.

Nevertheless, even if a detection of the Sun Lense-Thirring seems to be unreachable, the fact to not include this effect in the modelling of the orbital dynamics produce biases in the estimation of the other parameters. Our simulations indicate that not including the Lense-Thirring in the equations of motion leads to a bias at the level of $10^{-8}$ on $J_2$ and at the level of $5\times 10^{-5}$ on the $\beta$ PPN parameter. Similar conclusions seem to hold for planetary ephemerides, which is also mentioned by \citet{iorio:2011ys} and \citet{folkner:2014uq} for the $J_2$.

\subsection{Fifth force formalism}
The fifth force formalism consists in a Yukawa modification of the Newtonian potential~\citep{talmadge:1988uq,adelberger:2009fk}. This Yukawa modification is parametrized by a range of interaction $\lambda$ and by an intensity $\alpha$. Constraints on these two parameters can be found for example in Fig.~31 of~\citep{konopliv:2011dq}. At astronomical scales, the constraints are provided mainly by Lunar Laser Ranging and by ranging to Mars spacecraft. With simulation of 10,000 asteroids, we estimate the uncertainty on $\alpha$ for different fixed values of $\lambda$. At the level of $\lambda=10^{10}$ m, the uncertainty on $\alpha$ is slightly larger than the one from ranging to Mars spacecraft while for $\lambda=10^{11}$ m, our estimated uncertainty improves the current constraint by a factor 5. Therefore, this seems to be a very promising test. The correlation with the Sun mass needs nevertheless to be assessed (see the related discussion by \citet{konopliv:2011dq}).

\subsection{Temporal variation of the gravitational constant}
A lot of alternative theories of gravitation promotes the gravitation constant $G$ to a dynamical field (typically to a scalar field). In this case, $G$ becomes space-time dependant and can evolve for example with the cosmological evolution. Therefore, this class of models predicts a non-vanishing value of $\dot G$ in the Solar System. Our covariance study of the observations of 10,000 asteroids over 5 years indicates that Gaia will be able to constrain a linear evolution of the gravitational constant $\dot G/G$ at the level of $10^{-12}$ yr$^{-1}$ (or more precisely $d \ln GM_{\astrosun}/dt$). The correlation with the Sun $J_2$ is very weak (see also~\citep{mouret:2011uq}). The best current estimations on $\dot G/G$ are at the level of $10^{-13}$ yr$^{-1}$ and are provided by planetary ephemerides~\citep{konopliv:2011dq,fienga:2015rm}.

Moreover, very recently, it has been reported that the measurements of the gravitational constant $G$ seem to undergo some periodic variations~\citep{anderson:2015yq} with a 5.9 years period. This analysis has been confirmed by~\citet{schlamminger:2015rz} even if the conclusion has been weakened. We have performed a simulation including a periodic variation of $G$. This shows that asteroid observations from Gaia will be able to constrain the relative amplitude of the  oscillations at the level of $10^{-10}$. This accuracy is 5 orders of magnitude smaller than the amplitude predicted by \citet{anderson:2015yq,schlamminger:2015rz}. No correlation is expected with the Sun $J_2$. This means that Gaia will be able to rule out such a temporal variation of $G$. Note that, as mentioned by~\citet{iorio:2015zl}, the current planetary ephemerides analysis is also able to rule out this time variation with a similar accuracy.

\subsection{Standard Model Extension formalism}
The Standard Model Extension (SME) formalism has been developed to systematically describe violations of the Lorentz symmetry in all sectors of physics, including the gravitational sector. Consequences of SME on gravitational observations are developed into details in~\citep{bailey:2006uq,kostelecky:2011kx}. In the linearized gravity approximation, the gravitational sector of the minimal SME is described by a space-time metric that depends on a symmetry trace-free tensor $\bar s^{\mu\nu}$. It can be shown that the orbital dynamics is insensitive to $\bar s^{TT}$~\citep{bailey:2006uq} and therefore depends on 8 independent parameters: $\bar s^{XX}-\bar s^{YY}$, $\bar s^{XX}+\bar s^{YY}-2\bar s^{ZZ}$, $\bar s^{XY}$, $\bar s^{XZ}$, $\bar s^{YZ}$, $\bar s^{TX}$, $\bar s^{TY}$ and $\bar s^{TZ}$. The heliocentric equations of motion can be found in~\citep{bailey:2006uq}. 

The covariance analysis performed by considering 10,000 asteroids leads to the estimated uncertainties presented in Table~\ref{tab:SME}. These uncertainties are better than the current best estimations of the SME parameters available in the literature~\citep{kostelecky:2011ly}. In particular, they are better than the ones obtained with planetary ephemerides~\citep{iorio:2012zr,hees:2015sf}. This is due to the variety of the asteroids orbital parameters while planetary ephemerides use only 8 planets with similar orbital parameters (same orbital planes and nearly circular orbits). Therefore, the estimation of the SME parameters with planetary ephemerides are degraded by these correlations (see the discussion in~\citet{hees:2015sf}). Using our set of asteroids, the correlation matrix for the SME parameters is very reasonable: the three most important correlation coefficients are 0.71, -0.68 and 0.46. All the other correlations are below 0.3. Moreover, the Sun $J_2$ is not correlated to these parameters. Therefore, Gaia offers a unique opportunity to constrain Lorentz violation through the SME formalism. A combined analysis with planetary ephemerides analysis~\citep{hees:2015sf}, Lunar Laser Ranging~\citep{battat:2007uq} and atom interferometry~\citep{muller:2008kx} would also be very interesting. Our analysis needs to be extended to include gravity-matter Lorentz violation parametrized by $(\bar a)^\mu$ coefficients in the SME framework.

\begin{table}[hbt]
\caption{Sensitivity on the SME gravity parameters.}              % title of Table
\label{tab:SME}      % is used to refer this table in the text
\centering                                      % used for centering table
\begin{tabular}{c | c }          % centered columns (4 columns)
\hline\hline                        % inserts double horizontal lines
SME parameters & Sensitivity ($\sigma$)\\    % table heading
\hline                                   % inserts single horizontal line
     $\bar s^{XX}-\bar s^{YY}$  & $9\times 10^{-12}$ \\      % inserting body of the table
     $\bar s^{XX}+\bar s^{YY}-2\bar s^{ZZ}$ &  $2\times 10^{-11}$ \\
     $\bar s^{XY}$ & $4\times 10^{-12}$\\
     $\bar s^{XZ}$ & $2\times 10^{-12}$ \\
     $\bar s^{YZ}$ & $4\times 10^{-12}$ \\
     $\bar s^{TX}$ & $1\times 10^{-8\phantom{1}}$\\
     $\bar s^{TY}$ & $2\times 10^{-8\phantom{1}}$\\
     $\bar s^{TZ}$ & $4\times 10^{-8\phantom{1}}$ \\
\hline                                             %inserts single line
\end{tabular}
\end{table}

\section{Conclusion}
In this communication, we have presented different possibilities to use asteroid observations from Gaia to perform different test of the gravitation theory. The estimation of the $\beta$ PPN parameter and of a linear time dependance of the gravitational constant ($d \ln GM_{\astrosun}/dt$) would not be as good as the current estimations from planetary ephemerides. Nevertheless, the expected constraints are very complementary since they do not suffer from the same systematics errors.

Moreover, we have shown that the asteroid orbits are sensitive to a violation of the Strong Equivalence Principle through the $\eta$ parameter. In the framework of the PPN formalism, this $\eta$ parameter is related to the standard PPN parameters and this relation can help to reduce the correlation between the Sun $J_2$ and the PPN parameter $\beta$. Similar conclusion holds for planetary ephemerides analysis.

We have shown that the Sun Lense-Thirring effect is too weak to be detected with such dataset. Nevertheless, it would be interesting to see if a combination of the Gaia dataset with radar observations of asteroids~\citep{margot:2010fk} or with planetary ephemerides can lead to a first dynamical detection of the Sun angular momentum through the Lense-Thirring. Nevertheless, we have shown that not including the Sun Lense-Thirring in the equations of motion leads to a bias in the estimation of the Sun $J_2$ and of the $\beta$ PPN parameter. Similar conclusions hold for planetary ephemerides analysis.

The asteroid trajectories are also sensitive to a fifth force and can be used to improve constraints in this formalism.

Finally, the number of asteroids and the variety of their orbital parameters provide a unique opportunity to constrain Lorentz violation through the Standard Model Extension formalism. The wide orbital parameters allows to decorrelate the SME parameters and will allow us to produce the best estimations on the SME parameters.

Extended simulations considering the full set of asteroids (360,000 asteroids instead of 10,000 considered in this analysis) with refined astrometric precision (instead of 0.2 mas) and possible mission extension (6 years) are currently on-going. We are also currently assessing the improvement provided by combining the Gaia dataset with radar observations~\citep{margot:2010fk} that are complementary in the time frame and orthogonal to astrometric telescopic observations.

% Optional acknowledgements
% -------------------------
\begin{acknowledgements}
A.H. thanks the GRAM for financial support to attend this meeting and acknowledges support from ``Fonds Sp\'ecial de Recherche" through a FSR-UCL grant and from a GREAT-ESF exchange grant. A.H. thanks Q. G. Bailey for interesting discussions about SME and L. Iorio for his interesting comments. C.L.P.L. thanks GRAM and GPHYS from Paris Observatory for financial support.
\end{acknowledgements}

%%-----------------------------
%%   Bibliography
%%-----------------------------
%%

%% The following lines are required when using BibTEX (strongly encouraged!):
\bibliographystyle{aa}  % A&A bibliography style file (aa.bst)
\bibliography{../../../../Dropbox/JPL/JPL_byMe/biblio} % your references in file: Yourfile.bib

\begin{thebibliography}{51}
\expandafter\ifx\csname natexlab\endcsname\relax\def\natexlab#1{#1}\fi

\bibitem[{{Adelberger} {et~al.}(2009){Adelberger}, {Gundlach}, {Heckel},
  {Hoedl}, \& {Schlamminger}}]{adelberger:2009fk}
{Adelberger}, E.~G., {Gundlach}, J.~H., {Heckel}, B.~R., {Hoedl}, S., \&
  {Schlamminger}, S. 2009, \doi{10.1016/j.ppnp.2008.08.002}{Progress in Particle and Nuclear Physics, 62, 102}

\bibitem[{{Anderson} {et~al.}(2015){Anderson}, {Schubert}, {Trimble}, \&
  {Feldman}}]{anderson:2015yq}
{Anderson}, J.~D., {Schubert}, G., {Trimble}, V., \& {Feldman}, M.~R. 2015, \doi{10.1209/0295-5075/110/10002
}{EPL
  (Europhysics Letters), 110, 10002}

\bibitem[{Bailey \& Kosteleck\'y(2006)}]{bailey:2006uq}
Bailey, Q.~G. \& Kosteleck\'y, V.~A. 2006, \doi{10.1103/PhysRevD.74.045001}{Phys. Rev. D, 74, 045001}

\bibitem[{{Battat} {et~al.}(2007){Battat}, {Chandler}, \&
  {Stubbs}}]{battat:2007uq}
{Battat}, J.~B.~R., {Chandler}, J.~F., \& {Stubbs}, C.~W. 2007, \doi{10.1103/PhysRevLett.99.241103}{Physical Review
  Letters, 99, 241103}

\bibitem[{{Bertotti} {et~al.}(2003){Bertotti}, {Iess}, \&
  {Tortora}}]{bertotti:2003uq}
{Bertotti}, B., {Iess}, L., \& {Tortora}, P. 2003, \doi{10.1038/nature01997}{Nature, 425, 374}

\bibitem[{{Ciufolini} \& {Pavlis}(2004)}]{ciufolini:2004uq}
{Ciufolini}, I. \& {Pavlis}, E.~C. 2004, \doi{10.1038/nature03007}{\nat, 431, 958}

\bibitem[{{Delva} {et~al.}(2015){Delva}, {Hees}, {Bertone}, {Richard}, \&
  {Wolf}}]{delva:2015fk}
{Delva}, P., {Hees}, A., {Bertone}, S., {Richard}, E., \& {Wolf}, P. 2015,
  \doi{10.1088/0264-9381/32/23/232003}{Classical and Quantum Gravity, 32, 232003}


\bibitem[{{Everitt} {et~al.}(2011){Everitt}, {Debra}, {Parkinson}, {Turneaure},
  {Conklin}, {Heifetz}, {Keiser}, {Silbergleit}, {Holmes}, {Kolodziejczak},
  {Al-Meshari}, {Mester}, {Muhlfelder}, {Solomonik}, {Stahl}, {Worden},
  {Bencze}, {Buchman}, {Clarke}, {Al-Jadaan}, {Al-Jibreen}, {Li}, {Lipa},
  {Lockhart}, {Al-Suwaidan}, {Taber}, \& {Wang}}]{everitt:2011fk}
{Everitt}, C.~W.~F., {Debra}, D.~B., {Parkinson}, B.~W., {et~al.} 2011,
  \doi{10.1103/PhysRevLett.106.221101}{Physical Review Letters, 106, 221101}

\bibitem[{{Fienga} {et~al.}(2015){Fienga}, {Laskar}, {Exertier}, {Manche}, \&
  {Gastineau}}]{fienga:2015rm}
{Fienga}, A., {Laskar}, J., {Exertier}, P., {Manche}, H., \& {Gastineau}, M.
  2015, \doi{10.1007/s10569-015-9639-y}{Celestial Mechanics and Dynamical Astronomy, 123, 325}

\bibitem[{{Fienga} {et~al.}(2011){Fienga}, {Laskar}, {Kuchynka}, {Manche},
  {Desvignes}, {Gastineau}, {Cognard}, \& {Theureau}}]{fienga:2011qf}
{Fienga}, A., {Laskar}, J., {Kuchynka}, P., {et~al.} 2011, \doi{10.1007/s10569-011-9377-8}{Celestial Mechanics
  and Dynamical Astronomy, 111, 363}

\bibitem[{{Fienga} {et~al.}(2013){Fienga}, {Manche}, {Laskar}, {Gastineau}, \&
  {Verma}}]{fienga:2013yg}
{Fienga}, A., {Manche}, H., {Laskar}, J., {Gastineau}, M., \& {Verma}, A. 2013,
 \href{http://arxiv.org/abs/1301.1510}{arXiv:1301.1510}

\bibitem[{{Folkner} {et~al.}(2014){Folkner}, {Williams}, {Boggs}, {Park}, \&
  {Kuchynka}}]{folkner:2014uq}
{Folkner}, W.~M., {Williams}, J.~G., {Boggs}, D.~H., {Park}, R., \& {Kuchynka},
  P. 2014, \href{http://ipnpr.jpl.nasa.gov/progress_report/42-196/196C.pdf}{IPN Progress Report, 42}

\bibitem[{{Gu\'ena} {et~al.}(2012){Gu\'ena}, {Abgrall}, {Rovera}, {Rosenbusch},
  {Tobar}, {Laurent}, {Clairon}, \& {Bize}}]{guena:2012ys}
{Gu\'ena}, J., {Abgrall}, M., {Rovera}, D., {et~al.} 2012, \doi{10.1103/PhysRevLett.109.080801}{Physical Review
  Letters, 109, 080801}

\bibitem[{{Hees} {et~al.}(2015){Hees}, {Bailey}, {Le Poncin-Lafitte},
  {Bourgoin}, {Rivoldini}, {Lamine}, {Meynadier}, {Guerlin}, \&
  {Wolf}}]{hees:2015sf}
{Hees}, A., {Bailey}, Q.~G., {Le Poncin-Lafitte}, C., {et~al.} 2015, \doi{10.1103/PhysRevD.92.064049}{\prd, 92, 064049}

\bibitem[{{Hees} {et~al.}(2014{\natexlab{a}}){Hees}, {Bertone}, \& {Le
  Poncin-Lafitte}}]{hees:2014nr}
{Hees}, A., {Bertone}, S., \& {Le Poncin-Lafitte}, C. 2014{\natexlab{a}}, \doi{10.1103/PhysRevD.90.084020}{\prd,
  90, 084020}

\bibitem[{{Hees} {et~al.}(2014{\natexlab{b}}){Hees}, {Bertone}, \& {Le
  Poncin-Lafitte}}]{hees:2014fk}
{Hees}, A., {Bertone}, S., \& {Le Poncin-Lafitte}, C. 2014{\natexlab{b}}, \doi{10.1103/PhysRevD.89.064045}{\prd,
  89, 064045}

\bibitem[{{Hees} {et~al.}(2014{\natexlab{c}}){Hees}, {Folkner}, {Jacobson},
  {Park}, {Lamine}, {Le Poncin-Lafitte}, \& {Wolf}}]{hees:2013vn}
{Hees}, A., {Folkner}, W.~M., {Jacobson}, R.~A., {et~al.} 2014{\natexlab{c}},
  in {Proceedings of the Journ{\'e}es 2013 "Syst\`emes de R\'ef\'erence
  Spatio-Temporels"}, ed. N.~{Capitaine}, {Observatoire de Paris}
  ({Observatoire de Paris}), 241, \href{http://arxiv.org/abs/1403.1365}{arXiv:1403.1365}

\bibitem[{{Hees} {et~al.}(2014{\natexlab{d}}){Hees}, {Folkner}, {Jacobson}, \&
  {Park}}]{hees:2014jk}
{Hees}, A., {Folkner}, W.~M., {Jacobson}, R.~A., \& {Park}, R.~S.
  2014{\natexlab{d}}, \doi{10.1103/PhysRevD.89.102002}{\prd, 89, 102002}

\bibitem[{{Hees} {et~al.}(2012){Hees}, {Lamine}, {Reynaud}, {Jaekel}, {Le
  Poncin-Lafitte}, {Lainey}, {F{\"u}zfa}, {Courty}, {Dehant}, \&
  {Wolf}}]{hees:2012fk}
{Hees}, A., {Lamine}, B., {Reynaud}, S., {et~al.} 2012, \doi{10.1088/0264-9381/29/23/235027}{Classical and Quantum
  Gravity, 29, 235027}

\bibitem[{{Hees} {et~al.}(2014{\natexlab{e}}){Hees}, {Minazzoli}, \&
  {Larena}}]{hees:2014uq}
{Hees}, A., {Minazzoli}, O., \& {Larena}, J. 2014{\natexlab{e}}, \doi{10.1103/PhysRevD.90.124064}{Phys. Rev. D,
  90, 124064}

\bibitem[{{Hestroffer} {et~al.}(2010){Hestroffer}, {Mouret}, {Mignard},
  {Tanga}, \& {Berthier}}]{hestroffer:2010qv}
{Hestroffer}, D., {Mouret}, S., {Mignard}, F., {Tanga}, P., \& {Berthier}, J.
  2010, \doi{10.1017/S1743921309990585}{in IAU Symposium, Vol. 261, IAU Symposium, ed. S.~A. {Klioner}, P.~K.
  {Seidelmann}, \& M.~H. {Soffel}, 325--330}

\bibitem[{{Iorio}(2011)}]{iorio:2011zr}
{Iorio}, L. 2011, \doi{10.1016/j.asr.2011.06.023}{Advances in Space Research, 48, 1403}

\bibitem[{{Iorio}(2012{\natexlab{a}})}]{iorio:2012uq}
{Iorio}, L. 2012{\natexlab{a}}, \doi{10.1007/s11207-012-0086-6}{\solphys, 281, 815}

\bibitem[{{Iorio}(2012{\natexlab{b}})}]{iorio:2012fr}
{Iorio}, L. 2012{\natexlab{b}}, \doi{10.1007/s10714-011-1302-7}{General Relativity and Gravitation, 44, 719}

\bibitem[{{Iorio}(2012{\natexlab{c}})}]{iorio:2012zr}
{Iorio}, L. 2012{\natexlab{c}}, \doi{10.1007/s10509-010-0489-5}{Classical and Quantum Gravity, 29, 175007}

\bibitem[{{Iorio}(2015)}]{iorio:2015zl}
{Iorio}, L. 2015, \href{http://arxiv.org/abs/1504.07233}{arXiv:1504.07233}

\bibitem[{{Iorio} {et~al.}(2011){Iorio}, {Lichtenegger}, {Ruggiero}, \&
  {Corda}}]{iorio:2011ys}
{Iorio}, L., {Lichtenegger}, H.~I.~M., {Ruggiero}, M.~L., \& {Corda}, C. 2011,
  \doi{10.1007/s10509-010-0489-5}{\apss, 331, 351}

\bibitem[{Konopliv {et~al.}(2011)Konopliv, Asmar, Folkner, Karatekin, Nunes,
  Smrekar, Yoder, \& Zuber}]{konopliv:2011dq}
Konopliv, A.~S., Asmar, S.~W., Folkner, W.~M., {et~al.} 2011, \doi{10.1016/j.icarus.2010.10.004
}{Icarus, 211, 401}

\bibitem[{{Kosteleck{\'y}} \& {Russell}(2011)}]{kostelecky:2011ly}
{Kosteleck{\'y}}, V.~A. \& {Russell}, N. 2011, \doi{10.1103/RevModPhys.83.11}{Reviews of Modern Physics, 83,
  11}

\bibitem[{Kosteleck\'y \& Tasson(2011)}]{kostelecky:2011kx}
Kosteleck\'y, V.~A. \& Tasson, J.~D. 2011, \doi{10.1103/PhysRevD.83.016013}{Phys. Rev. D, 83, 016013}

\bibitem[{{Lambert} \& {Le Poncin-Lafitte}(2009)}]{lambert:2009bh}
{Lambert}, S.~B. \& {Le Poncin-Lafitte}, C. 2009, \doi{10.1051/0004-6361/200911714}{A\&A, 499, 331}

\bibitem[{{Margot} \& {Giorgini}(2010)}]{margot:2010fk}
{Margot}, J.-L. \& {Giorgini}, J.~D. 2010, \doi{10.1017/S1743921309990366}{in IAU Symposium, Vol. 261, IAU
  Symposium, ed. S.~A. {Klioner}, P.~K. {Seidelmann}, \& M.~H. {Soffel},
  183--188}

\bibitem[{{Mignard} \& {Klioner}(2010)}]{mignard:2010kx}
{Mignard}, F. \& {Klioner}, S.~A. 2010, \doi{10.1017/S174392130999055X}{in IAU Symposium, Vol. 261, IAU
  Symposium, ed. S.~A. {Klioner}, P.~K. {Seidelmann}, \& M.~H. {Soffel},
  306--314}

\bibitem[{{Minazzoli} \& {Hees}(2014)}]{minazzoli:2014xz}
{Minazzoli}, O. \& {Hees}, A. 2014, \doi{10.1103/PhysRevD.90.023017}{\prd, 90, 023017}

\bibitem[{{Mouret}(2011)}]{mouret:2011uq}
{Mouret}, S. 2011, \doi{10.1103/PhysRevD.84.122001}{\prd, 84, 122001}

\bibitem[{{Mouret} \& {Mignard}(2011)}]{mouret:2011qf}
{Mouret}, S. \& {Mignard}, F. 2011, \doi{10.1111/j.1365-2966.2010.18168.x}{\mnras, 413, 741}

\bibitem[{{M{\"u}ller} {et~al.}(2008){M{\"u}ller}, {Chiow}, {Herrmann}, {Chu},
  \& {Chung}}]{muller:2008kx}
{M{\"u}ller}, H., {Chiow}, S.-W., {Herrmann}, S., {Chu}, S., \& {Chung}, K.-Y.
  2008, \doi{10.1103/PhysRevLett.100.031101}{Physical Review Letters, 100, 031101}

\bibitem[{{M{\"u}ller} {et~al.}(2012){M{\"u}ller}, {Hofmann}, \&
  {Biskupek}}]{muller:2012ys}
{M{\"u}ller}, J., {Hofmann}, F., \& {Biskupek}, L. 2012, \doi{10.1088/0264-9381/29/18/184006}{Classical and Quantum
  Gravity, 29, 184006}

\bibitem[{{Pijpers}(2006)}]{pijpers:2006kx}
{Pijpers}, F.~P. 2006, {Methods in helio- and asteroseismology} (Imperial
  College Press)

\bibitem[{{Pitjeva} \& {Pitjev}(2013)}]{pitjeva:2013fk}
{Pitjeva}, E.~V. \& {Pitjev}, N.~P. 2013, \doi{10.1093/mnras/stt695}{MNRAS, 432, 3431}

\bibitem[{{Pitjeva} \& {Pitjev}(2014)}]{pitjeva:2014fj}
{Pitjeva}, E.~V. \& {Pitjev}, N.~P. 2014, \doi{10.1007/s10569-014-9569-0}{Celestial Mechanics and Dynamical
  Astronomy, 119, 237}

\bibitem[{{Rosenband} {et~al.}(2008){Rosenband}, {Hume}, {Schmidt}, {Chou},
  {Brusch}, {Lorini}, {Oskay}, {Drullinger}, {Fortier}, {Stalnaker}, {Diddams},
  {Swann}, {Newbury}, {Itano}, {Wineland}, \& {Bergquist}}]{rosenband:2008fk}
{Rosenband}, T., {Hume}, D.~B., {Schmidt}, P.~O., {et~al.} 2008, \doi{10.1126/science.1154622}{Science, 319,
  1808}

\bibitem[{{Schlamminger} {et~al.}(2008){Schlamminger}, {Choi}, {Wagner},
  {Gundlach}, \& {Adelberger}}]{schlamminger:2008zr}
{Schlamminger}, S., {Choi}, K.-Y., {Wagner}, T.~A., {Gundlach}, J.~H., \&
  {Adelberger}, E.~G. 2008, \doi{10.1103/PhysRevLett.100.041101}{Physical Review Letters, 100, 041101}

\bibitem[{{Schlamminger} {et~al.}(2015){Schlamminger}, {Gundlach}, \&
  {Newman}}]{schlamminger:2015rz}
{Schlamminger}, S., {Gundlach}, J.~H., \& {Newman}, R.~D. 2015, \doi{10.1103/PhysRevD.91.121101}{\prd, 91,
  121101}

\bibitem[{{Talmadge} {et~al.}(1988){Talmadge}, {Berthias}, {Hellings}, \&
  {Standish}}]{talmadge:1988uq}
{Talmadge}, C., {Berthias}, J.-P., {Hellings}, R.~W., \& {Standish}, E.~M.
  1988, \doi{10.1103/PhysRevLett.61.1159}{Physical Review Letters, 61, 1159}

\bibitem[{{Teyssandier} \& {Le Poncin-Lafitte}(2008)}]{teyssandier:2008nx}
{Teyssandier}, P. \& {Le Poncin-Lafitte}, C. 2008, \doi{10.1088/0264-9381/25/14/145020
}{Classical and Quantum Gravity, 25, 145020}

\bibitem[{{Uzan}(2011)}]{uzan:2011vn}
{Uzan}, J.-P. 2011, \doi{10.12942/lrr-2011-2}{Living Reviews in Relativity, 14, 2}

\bibitem[{{Verma} {et~al.}(2014){Verma}, {Fienga}, {Laskar}, {Manche}, \&
  {Gastineau}}]{verma:2014jk}
{Verma}, A.~K., {Fienga}, A., {Laskar}, J., {Manche}, H., \& {Gastineau}, M.
  2014, \doi{10.1051/0004-6361/201322124}{\aap, 561, A115}

\bibitem[{{Vessot} {et~al.}(1980){Vessot}, {Levine}, {Mattison}, {Blomberg},
  {Hoffman}, {Nystrom}, {Farrel}, {Decher}, {Eby}, \&
  {Baugher}}]{vessot:1980fk}
{Vessot}, R.~F.~C., {Levine}, M.~W., {Mattison}, E.~M., {et~al.} 1980, \doi{10.1103/PhysRevLett.45.2081}{Physical
  Review Letters, 45, 2081}


\bibitem[{{Will}(1993)}]{will:1993fk}
{Will}, C.~M. 1993, {Theory and Experiment in Gravitational Physics}, ed.
  {Will, C.~M.}

\bibitem[{{Will}(2014)}]{will:2014la}
{Will}, C.~M. 2014, \doi{10.12942/lrr-2014-4}{Living Reviews in Relativity, 17, 4}

\bibitem[{{Williams} {et~al.}(2009){Williams}, {Turyshev}, \&
  {Boggs}}]{williams:2009ys}
{Williams}, J.~G., {Turyshev}, S.~G., \& {Boggs}, D.~H. 2009, \doi{10.1142/S021827180901500X}{International
  Journal of Modern Physics D, 18, 1129}

\end{thebibliography}
%\bibliography{hees1}

%
\end{document}